# Investigation of Deuterium-Loaded Materials Subject to X-Ray Exposure

Theresa L. Benyo, Bruce M. Steinetz, and
Robert C. Hendricks
National Aeronautics and Space Administration
Glenn Research Center
Cleveland, Ohio 44135

Richard E. Martin
Cleveland State University
Cleveland, Ohio 44115

Lawrence P. Forsley
JWK Corporation
Annandale, Virginia 22003

Christopher C. Daniels
The University of Akron
Akron, Ohio 44325

Arnon Chait
National Aeronautics and Space Administration
Glenn Research Center
Cleveland, Ohio 44135

Vladimir Pines and Marianna Pines
PineSci Consulting
Avon Lake, Ohio 44012

Nicholas Penney
Ohio Aerospace Institute
Brook Park, Ohio 44142

Tracy R. Kamm and Michael D. Becks
Vantage Partners LLC
Brook Park, Ohio 44142

## Summary

Results are presented from an exploratory study involving x-ray irradiation of select deuterated materials. Titanium deuteride (TiD$_2$) plus deuterated polyethylene ([-CD$_2$-]$_n$; DPE), DPE alone, and for control, hydrogen-based polyethylene ([-CH$_2$-]$_n$; HPE) samples and nondeuterated titanium samples were exposed to x-ray irradiation. These samples were exposed to various energy levels from 65 to 280 kV with prescribed electron flux from 500 to 9000 µA impinging on a tungsten braking target, with total exposure times ranging from 55 to 280 min. Gamma activity was measured using a high-purity germanium (HPGe) detector, and for all samples no gamma activity above background was detected. Alpha and beta activities were measured using a gas proportional counter, and for select samples beta activity was measured with a liquid scintillator spectrometer. The majority of the deuterated materials subjected to the microfocus x-ray irradiation exhibited postexposure beta activity above background and several showed short-lived alpha activity. The HPE and nondeuterated titanium control samples exposed to the x-ray irradiation showed no postexposure alpha or beta activities above background. Several of the samples (SL10A, SL16, SL17A) showed beta activity above background with a greater than 4σ confidence level, months after exposure. Portions of SL10A, SL16, and SL17A samples were also scanned using a beta scintillator and found to have beta activity in the tritium energy band, continuing without noticeable decay for over 12 months. Beta scintillation investigation of as-received materials (before x-ray exposure) showed no beta activity in the tritium energy band, indicating the beta emitters were not in the starting materials.

## 1.0 Introduction and Motivation

Power systems for deep-space and planetary missions that cannot rely on solar power have for the most part exploited heat sources based on $^{238}$Pu. For instance, the Mars Science Laboratory's Radioisotope Thermoelectric Generator provides 2 kW of thermal power, which is then converted to electric power (Ref. 1). Limited available quantities and the expense associated with the production and safe handling of this isotope of plutonium have motivated the NASA Glenn Research Center to seek alternative sources of energy that could be maintained over a decade or more.

---

[1]This document reports preliminary findings intended to solicit comments and ideas from the technical community and is subject to revision as analysis proceeds.

[2]Trade names and trademarks are used in this report for identification only. Their usage does not constitute an official endorsement, either expressed or implied, by the National Aeronautics and Space Administration.

The ideal energy source would be light and compact, maintenance free, would deliver gigajoule levels of energy over a decade in operation, would not rely on enriched fuel, and could be actively controlled. To date, nuclear-based power generation is the only known technology with the required power or energy density (power or energy per unit mass) that could continuously operate for an extended period. The Advanced Energy Conversion (AEC) effort at NASA Glenn is exploring alternative power sources that preferably subscribe to the above set of attributes.

Deuterium, an isotope of hydrogen with one proton and one neutron, has been used as a nuclear material for many decades for applications ranging from inertial confinement fusion (ICF) reactors through neutron generators. Deuterium owes some of its key properties to its $Z = 1$ nuclear positive charge and therefore possesses the lowest barrier for tunneling the electrostatic barrier for nuclear fusion. Deuterium fusion, however, generally requires at least 10 to 15 keV in kinetic energy (corresponding to over 100 million degrees Kelvin) to raise the probability of tunneling to occur. Although the subject of intense work over many decades, no "hot fusion" nuclear reactor with a coefficient of performance greater than 1 (net positive power output) has been demonstrated to date. Deuterium is also a stable isotope of hydrogen, meaning that it poses no danger to the environment via launch risk. Deuterium is available in nature, and it is separated from seawater, where it has natural abundance of 0.0156%. From a nuclear standpoint, it has the lowest binding energy of any isotope (although still quite significant at 2.225 MeV) (Ref. 2). Indeed, deuterium has been the focus of attention of many attempts over the years to exploit one or more of its unique properties to achieve alternative forms of nuclear activity.

If a novel nuclear reaction were to occur, it must follow conventional rate relations that describe all nuclear processes. Essentially, the rate of nuclear processes is proportional to the product of the respective number densities of the reactants, as well as other parameters (e.g., the Gamow factor describing the probability of two particles to overcome the electrostatic repulsion barrier in nuclear fusion). Therefore, for deuterium to participate in a reaction, it would be advantageous to bring its number density to near solid-state condition. In nature, such conditions are possible using deuterated metals, where the atomic ratio of deuterium to the host metal can be greater than unity under certain conditions. Moreover, many metals can be deuterated (i.e. loaded with deuterium) and do maintain such stoichiometry with easily accessible pressure and temperature conditions. There are also many materials, including organics, in which the hydrogen could be replaced with deuterium using conventional chemical means. For example, in the past, ICF targets have used deuterated polyethylene (DPE) (Ref. 3) as a source of deuterium at near-solid-deuterium densities.

The present study explores possible reactions that combine two key elements: deuterium at high number density and moderate-energy photons that create moderate-energy electrons (in the keV range). In recognition of the fundamental nuclear rate equation, the high-number-density deuterium is provided in the present experiments either embedded in deuterated metals or directly replacing hydrogen in a high-number-density organic polymeric material. The high-energy electrons were provided to the process via the photoelectron or Compton effects under irradiation of a microfocus x-ray beam source. Evidence of nuclear reactions was investigated by comparing pretest and posttest nuclear activity in the form of alpha and beta emissions.

## 2.0 Materials

Several means of achieving high-number-density ($>10^{21}$ atoms/cm$^3$) deuterium were considered that included high gas pressures (2000 to 3000 bar), deuterated metals, and both solid and liquid deuterated polymers. Materials used in the present set of experiments were titanium deuteride (TiD$_2$), DPE, and various mixtures thereof. Mixtures for the DPE and TiD$_2$ were either a uniform mixture (DPE+TiD$_2$) or alternating layers of DPE and TiD$_2$ (DPE/TiD$_2$). NASA Glenn obtained these materials from commercial suppliers.

## 2.1 Titanium Deuteride

Titanium can be loaded to greater than a 2:1 atomic ratio of hydrogen (Ref. 4), which is generally a good guide for estimating deuterium loading. When loaded to a 2:1 loading ratio, TiD$_2$ has a deuterium atomic number density of $1\times10^{23}$ atom/cm$^3$. For the x-ray set of experiments, deuterated titanium was used with a loading ratio of 1.9 to 2:1, as confirmed by mass gain during the deuteriding process. Another attractive feature of TiD$_2$ is that once loaded, it can be handled under ambient conditions without losing its high loading, making transferring from a loading apparatus to test apparatus much easier. The experiment was run with several sizes of TiD$_2$ particles that ranged from

(a) 1000 to 2000 μm
(b) 300 to 1000 μm
(c) 150 to 300 μm
(d) < 50 μm (Some TiD$_2$ materials were quite fine and appeared as almost black dust.)

## 2.2 Deuterated Polyethylene

Another key material investigated in these of experiments was DPE. Its structure is polymerized carbon (C) chains of CD$_2$, or more specifically [-CD$_2$-]$_n$. Fully deuterated DPE can have very high deuterium number density ($7\times10^{22}$ atoms/cm$^3$) and may give a practical path forward as a compact fuel. To achieve high number density, materials were purchased with nominal reference values of about 98% deuterated.

Results from the x-ray exposure of deuterated materials revealed varying levels of nuclear activity depending on the DPE used. Because of this, the DPE materials were analyzed using Fourier transform infrared (FTIR) spectroscopy to verify the level of deuteration. FTIR spectroscopy revealed the materials exhibited incomplete replacement of hydrogen (H) with deuterium (D) and the actual ratio of C-D (2800 cm$^{-1}$, doublet) to C-H stretching (2150 cm$^{-1}$, doublet) found in the DPE were in the ratio of ~1.5:1 to 2:1 (corresponding to 60% to 67% D-loading, respectively), instead of the expected ~50:1 (corresponding to 98% D-loading) as claimed by the manufacturer. Subsequent purchases from the same manufacturer showed an increase in D-loading to 97%. The DPE morphology varied from hard-cloth-like to sponge-like to puff-ball-like, between material lots and within the same 1-g manufacturer-supplied bottles. This variation in the ratio of C-D to C-H bonds could indicate that quality control varies within a single manufacturer from batch to batch.

## 2.3 Control Materials

Polyethylene is a thermoplastic polymer consisting of long hydrocarbon chains [-CH$_2$-]$_n$. Hydrogen-based ultra-high-molecular-weight polyethylene (HPE) was used as control, lacking D as the nuclear fuel. Nondeuterated titanium powder was also used as a control material.

## 3.0 Experiment Description

The goal of the present experiments was to explore a possible connection between x-ray-induced photoelectric and/or Compton scattered electrons and nuclear reactions in deuterated materials versus that in hydrogenated materials. The protocol generally included the following steps:

(1) Create documentation packages, called travelers, for tracking sample history for analysis and testing.

(2) Perform pretest materials characterization, including alpha and beta measurements, using a gas proportional counter (see Sec. 3.1), and in some cases, also using a beta scintillation spectrometer.
(3) Load the reactants into a fixture and evacuate. If desired in some cases, backfill with deuterium gas ($\leq$350 psia, 2.4 MPa).
(4) Align the fixture into the beam of an x-ray.
(5) Expose fixture to the x-ray beam per experimental protocol.
(6) Unload the reactants.
(7) Perform posttest materials characterization. Perform alpha and beta measurements using the gas flow proportional counter and in some cases, also using a beta scintillation spectrometer. After alpha and beta measurements using the gas flow proportional counter, perform gamma measurements using a high-purity germanium detector (HPGe).
(8) Compile data into the travelers as well as electronically.
(9) Store the samples for archive purposes or for repeated activity measurements.

## 3.1 Test Fixture

During each experiment the materials were loaded into a 316 stainless steel Swagelok test fixture. An example test fixture is shown in Figure 1. During the experiment, key operational parameters were monitored such as internal and external temperatures and internal pressure. Note: there were no significant temperature or pressure changes observed that could not be explained by factors related to the experimental apparatus including natural x-ray tube heating with extended beam runtimes.

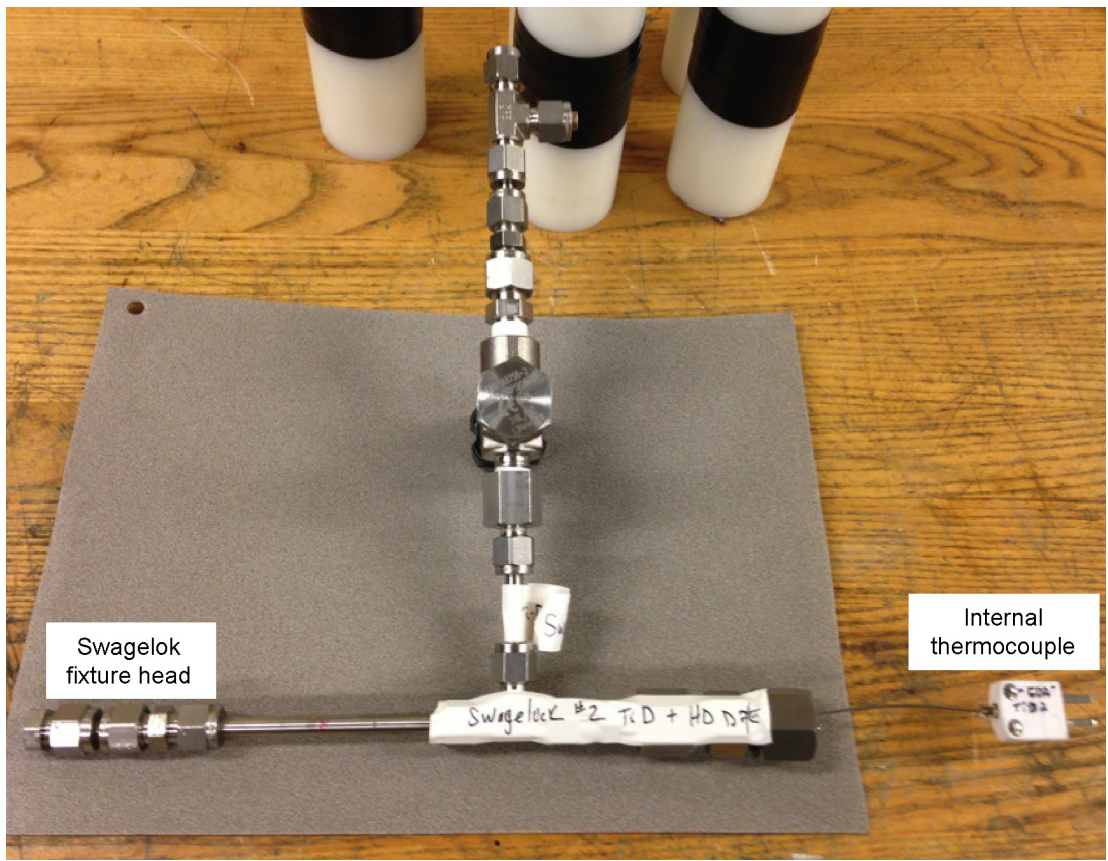

Figure 1.—Swagelok test fixture with thermocouple.

The interchangeable fixture consisted of a plug, a connecting tube, a tube-to-tube adaptor, and a screen to retain particles and to center the internal thermocouple. The Swagelok "heads" used in these experiments were 1/4, 3/8, and 1/2 in., based on tube size and experimental needs. The tubes were made of heavy-wall stainless steel. All elements were initially cleaned with acetone and alcohol, and subsequently each head component was cleaned with acetone and/or alcohol or both prior to each test material loading. The majority of the tests started with the Swagelok tube evacuated. Only a few test fixtures were backfilled with deuterium gas, including SL18 (350 psia) and SL25 (ambient deuterium).

## 3.2 Test Fixture Loading

Each test material was loaded into the Swagelok head using a variety of configurations. For DPE only, they were simply loaded and tamped into place using precleaned small-diameter rods. The DPE and $TiD_2$ particles were loaded in one of two ways: (1) In the "mixed" DPE+$TiD_2$ samples, the materials were combined and mixed by stirring with a rod in a cleaned container. These were then fed into the Swagelok fixture head. The weight used was determined by differences in the mixing container weight. (2) In the "layered" DPE/$TiD_2$ samples, alternating layers of DPE and $TiD_2$ particles were deposited, until the tube head was filled. Figure 2 shows an example of "layered" loading with a Swagelok fixture head. To maximize the loading they were packed a "pinch" at a time and tamped into place with the small-diameter rod.

After the tube head was filled, the plug was then secured, and the Swagelok was prepared for an argon pressure leak check at 1 MPa (145 psia) and subsequently vacuum leak checked by pulling a vacuum and determining if the pressure would rise. Where practical, the fixture was packed the night before and held under vacuum. If the pressure was found to be rising from vacuum, then one of the following actions were taken. If it were a vacuum test, a cover gas of argon was kept on the fixture overnight (rather than allowing room air to leak back into the chamber). If the test was to be run with deuterium gas, then a slight pressure of deuterium gas was used as a cover gas until the x-ray test was run. It was desired to prevent long periods of the reactants sitting under an air-moisture environment before testing.

## 3.3 Material Matrix and Test Plan

A total of 26 test samples were exposed in this study with various configurations of material and exposed to varying x-ray beam strengths for a variety of total exposure times. Table I shows the material contents of each Swagelok test fixture.

### 3.3.1 Tests With High-Number-Density Deuterium Materials

*DPE only*.—Swagelok fixtures SL1 and SL3 were loaded with DPE only, to determine if just the presence of high-number-density polymers could result in photon-induced activity. These fixtures were packed into the fixture using the methods noted above.

*DPE and $TiD_2$*.—Swagelok fixtures SL2, SL4, SL10, SL10A, SL15, SL15A, SL16, SL17–1, SL17A, SL18, SL18A, SL18B, SL19, SL20, SL23, SL25, and SL26 were loaded with both DPE and $TiD_2$, as either a mixture (DPE+$TiD_2$) or layered (DPE/$TiD_2$) as noted in Table I. Table II provides the measured alpha and beta activity in disintegrations per minute (dpm) of as-received $TiD_2$ or DPE materials before x-ray exposure, measured using the Tennelec alpha and beta counter. None of the as-received (i.e., pretest) materials showed activity above the minimum detectable amount (MDA). MDA is further explained in Appendix A.

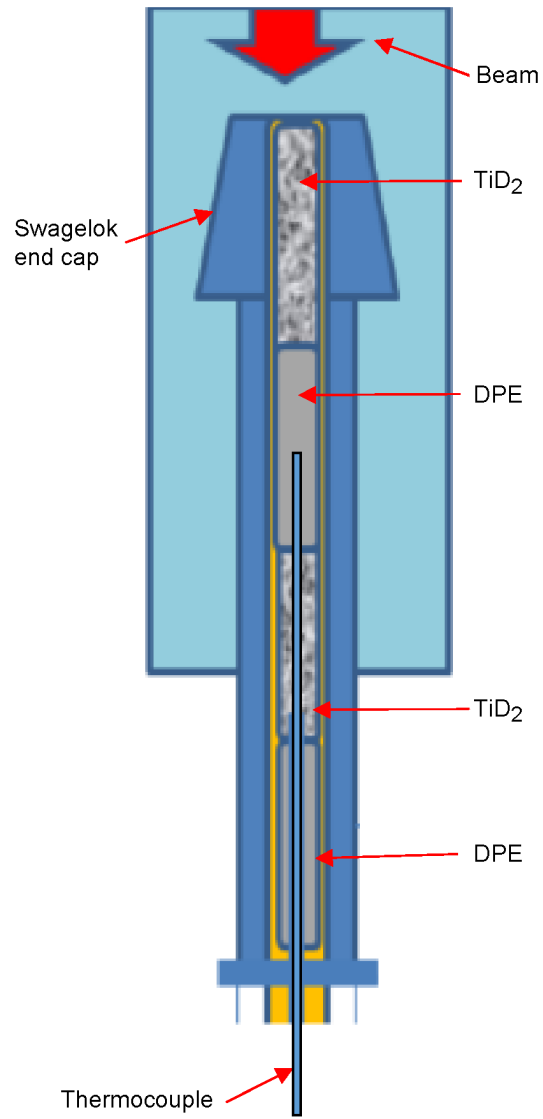

Figure 2.—Swagelok test fixture with layers of titanium deuteride (TiD$_2$) and deuterated polyethylene (DPE) loaded into fixture head.

TABLE I.—MATERIAL TEST MATRIX

| Sample | Material contents[a] | | | | Configuration | Sample | Material contents[a] | | | | Configuration |
| --- | --- | --- | --- | --- | --- | --- | --- | --- | --- | --- | --- |
| | Material A | | Material B | | | | Material A | | Material B | | |
| | Material | Mass, g | Material | Mass, g | | | Material | Mass, g | Material | Mass, g | |
| SL1 | DPE | 0.083 | NA | --- | S | SL16 | DPE | 0.594 | $TiD_2$ | 0.454 | M |
| SL2 | DPE/$TiD_2$ Total = 0.17 | | --- | | L | SL17–1 | DPE | 0.16 | $TiD_2$ | 1.133 | L |
| | | | | | | SL17A | DPE | 0.275 | $TiD_2$ | 1.165 | L |
| SL3 | DPE | 0.479 | NA | --- | S | SL18 | DPE | 0.17 | $TiD_2$ | 1.17 | L |
| SL4 | DPE | 0.154 | $TiD_2$ | 1.748 | L | SL18A | DPE | 0.17 | $TiD_2$ | 1.17 | L |
| SL6 | HPE | 0.312 | NA | --- | S | SL18B | DPE | 0.17 | $TiD_2$ | 1.17 | L |
| SL8 | HPE | 0.594 | NA | --- | S | SL19 | DPE | 0.536 | $TiD_2$ | 4.58 | L |
| SL10 | DPE+$TiD_2$ Total = 0.761 | | --- | | M | SL20 | DPE | 0.449 | $TiD_2$ | 0.87 | M |
| | | | | | | SL21 | HPE | 0.83 | NA | --- | S |
| SL10A | DPE+$TiD_2$ Total = 0.761 | | --- | | M | SL23 | DPE | 0.518 | $TiD_2$ | 0.391 | L |
| | | | | | | SL24 | HPE | 1.072 | NA | --- | S |
| SL14 | HPE | 0.83 | NA | --- | S | SL25 | DPE | 0.521 | $TiD_2$ | 1.436 | |
| SL15 | DPE | 0.57 | $TiD_2$ | 0.89 | M | SL26 | DPE | 0.645 | $TiD_2$ | 0.81 | L |
| SL15A | DPE | 0.57 | $TiD_2$ | 0.89 | M | | | | | | |
| Null tests | | | | | | | | | | | |
| SL61 (vacuum) | DPE | 0.552 | $TiD_2$ | 0.363 | L | SL65 (vacuum) | Ti | 2 | NA | --- | S |

[a]S is single material, L is layer, and M is mixture.

TABLE II.—ALPHA AND BETA ACTIVITY OF AS-RECEIVED $TiD_2$ OR DEUTERATED POLYETHYLENE (DPE) MATERIALS BEFORE X-RAY EXPOSURE[a]

[Negative activity entries are reported as zero. See Appendix A for more details.]

| Sample | | Alpha | | | Beta | | |
| --- | --- | --- | --- | --- | --- | --- | --- |
| ID number | Material | Alpha, dpm | Uncertainty, dpm | MDA,[b] dpm | Beta, dpm | Uncertainty, dpm | MDA,[b] dpm |
| AEC-Ti-GL-D-20140915-001 | $TiD_2$ | 0 | 0.32 | 2.08 | 0 | 2.05 | 4.78 |
| AEC-DPE-NL-D-20140904-1 | DPE | 1.27 | 0.83 | 2.08 | 0 | 2.48 | 4.78 |
| AEC-DPE-NL-D-20140915-001 | DPE | 0.21 | 0.65 | 2.54 | 0.34 | 3.42 | 6.93 |
| AEC-DPE-NL-D-20140915-002 (1/2 of total:A) | DPE | 0.21 | 0.65 | 2.54 | 0 | 3.16 | 6.93 |
| AEC-DPE-NL-D-20140915-002 (1/2 of total:B) | DPE | 0.21 | 0.65 | 2.54 | 0 | 3.38 | 6.93 |
| DPE 0341576 | DPE | 0.61 | 0.98 | 2.9 | 0 | 3.22 | 7.23 |
| DPE 0341634 | DPE | 0.1 | 0.83 | 2.9 | 0 | 3.22 | 7.23 |
| DPE 0341636 | DPE | 0.61 | 0.98 | 2.9 | 2.19 | 3.51 | 7.23 |
| DPE 0341637 | DPE | 1.63 | 1.21 | 2.9 | 4.84 | 3.85 | 7.23 |
| DPE 0343608 | DPE | 0.35 | 0.76 | 2.35 | 2.6 | 4.66 | 5.82 |

[a]Measured in disintegrations per minute (dpm).

[b]MDA is minimum detectable amount.

TABLE III.—ALPHA AND BETA ACTIVITY OF AS-RECEIVED HYDROGEN-BASED POLYETHYLENE (HPE) MATERIALS BEFORE X-RAY EXPOSURE[a]

[Negative activity entries are reported as zero. See Appendix A for more details.]

| Sample | | Alpha | | | Beta | | |
|---|---|---|---|---|---|---|---|
| ID number | Material | Alpha, dpm | Uncertainty, dpm | MDA,[b] dpm | Beta, dpm | Uncertainty, dpm | MDA,[b] dpm |
| AEC-HPE-NL-H-20140916-001 | HPE | 0.12 | 0.5 | 2.08 | 0 | 2.24 | 4.78 |
| AEC-HPE-NL-H-20141221-001 | HPE | 0 | 0.66 | 2.9 | 3.78 | 3.47 | 7.23 |
| AEC-HPE-NL-H-20141223-001 | HPE | 0.35 | 0.76 | 2.35 | 0 | 4.48 | 5.82 |
| AEC-HPE-NL-H-20141223-002 | HPE | 0 | 0.66 | 2.9 | 0 | 3.13 | 7.23 |

[a]Measured in disintegrations per minute (dpm).
[a]MDA is minimum detectable amount.

### 3.3.2 Control Tests

Several tests were conducted to better understand the parameter space of elements leading to activation or no activation. These control tests help identify the significance of this element in contributing to activation. Separate tests were conducted using HPE with x-ray exposure (no deuterium loading); some were conducted using deuterated materials without x-ray exposure, and one was conducted using titanium powder (no deuterium loading) with x-ray exposure.

*HPE materials*.—Five tests (SL6, SL8, SL14, SL21, and SL24) were run with HPE to serve as a control against which to compare any activation findings from the deuterated targets. Table III provides the measured alpha and beta activity of as-received HPE materials before x-ray exposure, measured using the Tennelec alpha and beta counter. None of the materials showed activity above the MDA.

*Null test: no x-ray exposure*.—SL61 was built with DPE and $TiD_2$ materials per the layering procedure noted previously (Sec. 3.2). SL61 was held under vacuum. The fixture was set in front of the x-ray beam, but the beam was left off. This was to determine if there could have been any contamination (e.g., radon) or anything introduced during the loading that could have possibly resulted in accidental alpha and beta activation of the fixture contents.

*Null test: no fuel, with x-ray exposure*.—SL65 was created to investigate the hypothesis that simply exposing the materials to the ionizing beam of x-rays would cause the materials to become alpha and beta active. SL65 was loaded with standard titanium without deuterium loading. This fixture was placed in front of the beam for the times indicated in Table IV to determine the material response.

### 3.4 X-Ray Beam Instrument

The filled test fixtures described above were exposed to x-ray energy using commercially available x-ray sources. The majority of exposures for this study were performed using a microfocus x-ray system, Model XWT–225–SE, manufactured by X-RAY Worx GmbH and shown in Figure 3. The tube utilizes a tungsten target for x-ray generation and is capable of voltages up to 225 kV and currents up to 1 mA. The microfocus projects the electron beam on a small 10-μm spot size, which leads to a very high unit flux leaving the braking target of $1.3 \times 10^{-2}$ mA/μm$^2$. (Even though the microfocus could reach 225 kV, the beam current was less at that condition so tests were only conducted up to 200 kV when using this system). X-ray energy filters were not used for any tests. Figure 4 presents a Monte-Carlo based prediction of photon fluence versus applied electron energy potential. Results were based on materials and geometry from the manufacturer and were calculated using the EGSnrc code (Ref. 5).

TABLE IV.—X-RAY EXPOSURE PARAMETERS FOR EACH SAMPLE

[All exposures performed with microfocus beam with the exception as noted.]

| Sample | X-ray exposure parameters | | | Sample | X-ray exposure parameters | | |
|---|---|---|---|---|---|---|---|
| | Voltage, kV | Current, μA | Time, min | | Voltage, kV | Current, μA | Time, min |
| SL1<br>SL4 | 65<br>120<br>90<br>200 | 500<br>500<br>500<br>500 | 15<br>5<br>5<br>30 | SL17–1 | 65<br>120<br>200<br>200 | 500<br>500<br>500<br>1000 | 5<br>5<br>20<br>60 |
| SL2 | 65<br>120<br>200<br>200 | 500<br>500<br>500<br>1000 | 5<br>5<br>20<br>60 | SL18 | 65<br>120<br>200<br>200 | 500<br>500<br>500<br>1000 | 5<br>5<br>25<br>60 |
| SL3 | 120<br>200<br>200 | 500<br>500<br>1000 | 5<br>20<br>40 | SL18A (i.e., SL30) Muller beam | 65<br>120<br>200<br>200<br>240<br>280<br>120<br>65 | 2000<br>2000<br>2000<br>9000<br>9000<br>9000<br>9000<br>9000 | 20<br>20<br>30<br>60<br>60<br>60<br>15<br>15 |
| SL5<br>Vacuum test used to check instrumentation | 65<br>120<br>200<br>200 | 500<br>500<br>500<br>1000 | 5<br>5<br>20<br>60 | SL18B (i.e., SL32)<br>SL25<br>SL26 | 65<br>120<br>200<br>200 | 500<br>500<br>500<br>1000 | 20<br>20<br>20<br>90 |
| SL6<br>SL8<br>SL10 | 65<br>120<br>200<br>200 | 500<br>500<br>500<br>1000 | 5<br>5<br>20<br>60 | SL19 | 200<br>100 | 1000<br>2500 | 30<br>30 |
| SL10A | 65<br>120<br>200<br>200 | 500<br>500<br>500<br>1000 | 5<br>5<br>5<br>120 | SL20 | 65<br>120<br>200<br>200 | 500<br>500<br>500<br>1000 | 5<br>5<br>60<br>90 |
| SL14<br>SL15<br>SL16 | 65<br>120<br>200<br>200 | 500<br>500<br>500<br>1000 | 20<br>20<br>20<br>60 | SL21<br>SL23<br>SL24 | 65<br>120<br>200 | 500<br>500<br>500 | 5<br>5<br>5 |
| SL15A<br>SL17A | 65<br>120<br>200<br>200 | 500<br>500<br>500<br>1000 | 20<br>20<br>20<br>90 | | | | |
| Null tests | | | | | | | |
| SL61 | No x-ray exposure | | | SL65 | 65<br>120<br>180<br>180 | 500<br>500<br>500<br>1200 | 10<br>10<br>10<br>90 |

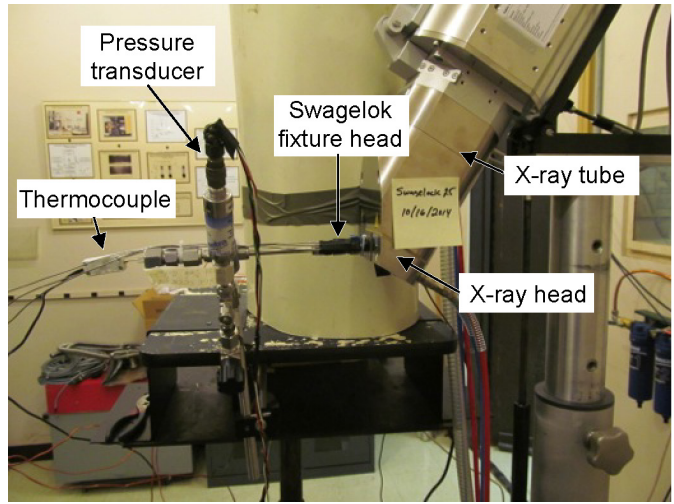

Figure 3.—Swagelok test fixture mounted in front of microfocus x-ray tube for exposure.

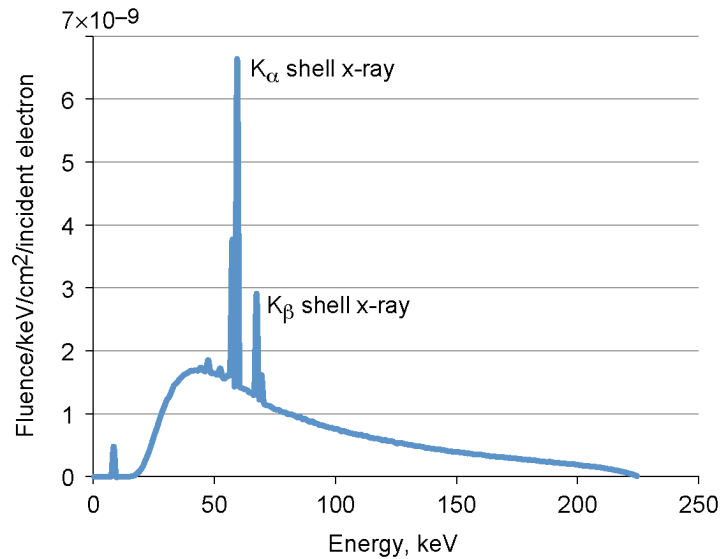

Figure 4.—Calculated microfocus photon fluence emission versus electron accelerating voltage. Note highest peak fluence is at tungsten $K_\alpha$ shell.

A limited number of tests were also conducted for comparison using a higher kilovolt energy Muller system Model No. 301/4 with settings of 280 kV and currents up to 9 mA. This system projects the electron beam on a larger 4-mm spot size, which leads to a significantly lower unit flux leaving the braking target of $7\times10^{-7}$ mA/µm$^2$. This system was only used to re-expose sample SL18A as will be discussed in the Experimental Results section. Beam time on this system was limited, preventing a comparable set of tests using this system, but the limited tests performed showed less success rate in beta activation than when using the microfocus beam.

### 3.5 Test Conditions

For tests conducted in the microfocus beam, x-rays are created at the tungsten-braking target positioned ~6 mm from the exit plane of the x-ray head. (Fig. 3). The Swagelok test fixtures were placed in close proximity (~1 mm) to the beam exit plane in order to maximize the intensity of the beam entering the sample. The distance from the braking target in the 280-kV head to the face of the Swagelok test fixture was 10 cm because of constraints imposed by the equipment. Tests were conducted using a ramp-up in both x-ray voltage and current. This ramp-up was done as a warmup for the tube to avoid damaging the system at higher power levels. Over the course of testing, some of the profile settings were changed. Table IV outlines the exposure energies and times for each of the samples used in this study.

## 3.6  Nuclear Activity Measurements

Nuclear activity of the samples in terms of alpha, beta, and gamma decay was measured with three types of nuclear diagnostic equipment.

### 3.6.1  Gamma Detection

Three separate HPGe detection systems were utilized to count the samples. All were Ortec units employing cryocoolers, using cylindrical lead caves with passive graded shielding made of tin and copper to reach low background counts, generally less than 25 counts per second (cps). The cylindrical lead cave around the HPGe crystal has a 6.35-cm lid thickness, 6-cm wall thickness, 15.24-cm inner diameter, 27.3-cm outer diameter, and 21-cm chamber height. Units 1 (Mod. No. MX45P–A–S) and 2 (Mod. No. GMX45P4–A) utilized an aluminum window, allowing photon energies to be measured down to about 40 keV. Unit 3 (Mod GMX40P4) utilized a beryllium window, allowing x-ray lines to be measured down to 25 keV.

Quality controls included daily spectral line checks with a $^{137}$Cs-, $^{152}$Eu-, and $^{241}$Am-certified check source. The check source has a total activity of about 18.5 kBq (0.5 µCurie) and known individual isotope activities as of the check source's issue date. Daily checks with the source include ensuring that the spectral lines of $^{137}$Cs, $^{152}$Eu, and $^{241}$Am are at the proper position and that the detected activity (kBq) is as expected in accordance with the known activities of each isotope. The detector's minimal detectable activity is on the order of 5.55 Bq (0.15 nCi). Along with daily source checks, daily background checks with an empty, closed cave are performed to ensure that the lead cave is not contaminated. The daily check ensures the calibrated energy and efficiency for each detector are correct with an occasional small gain change as necessary to maintain specifications using check sources. NASA Glenn personnel perform periodic calibrations on the order of once per year or more often if needed. All units were within their calibration specifications (calibration date Sept. 24, 2014). The manufacturer requires no calibration unless a detector cannot be brought into proper operating conditions. The spectra were displayed with GammaVision® 7 software (Ref. 6) using the NPP32 analysis engine.

### 3.6.2  Alpha and Beta Detection

As-received and postexposed materials were scanned with a Canberra (Tennelec) Series 5 XLB – Automatic Low Background Alpha/Beta Counting System gas flow proportional counter. This gas flow proportional counter was checked daily using plated alpha ($^{239}$Pu) and beta ($^{99}$Tc) sources as well as an empty planchet to ensure proper system performance and acceptable background levels.

As materials were received from manufacturers, a sample was split off for pre-exposed alpha and beta activity measurement. After it was confirmed there was no activity above background, portions of these as-received materials were then weighed, loaded into a Swagelok fixture (Fig. 1), and placed in the path of the beam. After the exposure had been completed, the contents were moved to the analysis lab, the fixture was opened, and the contents were emptied into an empty metal pan (blank planchet) and counted. Lack of proximity between the x-ray and analysis laboratories prevented collection of activity data within the first 30 min following completion of experiments. Hence, very short half-life elements would not be measured by the Tennelec alpha and beta counter. Appendix A provides additional details about the counter including the algorithms used to determine the low-count alpha and beta activity, assessing the MDA, measurement uncertainty, and other parameters of interest.

### 3.6.3 Liquid Scintillation

Two separate Beckman Coulter LS–6500 Liquid Scintillation Systems were used to assess beta activity in selected samples pretest and posttest. The NASA Glenn LS–6500 located in Cleveland utilized EcoScint fluid. The external partner's LS–6500 located in San Diego used Scintiverse fluid. A small sample (~ 0.5g) is immersed into a vial containing enough scintillation fluid (either EcoScint or Scintiverse) to completely cover the sample. The sample is then ready to be scanned with the liquid scintillation system.

The Beckman LS–6500 is used for biochemical radioisotope tracking, environmental sampling, or Radiation Safety Office "swipes." Generally, these involve known, or suspected, radioisotopes. In the earlier days of radiotracers some of the most common radiotracers were tritium, $^{14}$C, and $^{32}$P. Since the betas from these three radioisotopes are relatively distinct from each other, the LS–6500 bins the energies into these three bands. This "banding" is very useful in identifying the presence or absence of a beta emitter, but does not imply that particular isotope's presence; just energies in that band. The scintillator counting uncertainty with samples with low activity and 60-min counting times is ±5%.

The LS6500 system was checked daily using a $^{14}$C check source according to Beckman specifications to confirm that the instrument was properly counting the corresponding standard. Whenever a sample was counted, a background count was performed for reference purposes.

## 4.0 Experimental Results

After each x-ray exposure test was completed, the Swagelok fixture was removed from the x-ray head and transported to a fume hood where the fixture was taken apart and the material in the head was poured into a cleaned sample planchet. No obvious visual changes were noted in any sample as a results of x-ray exposure. The sample and planchet were then placed into the Tennelec for alpha and beta counting. Figure 5 shows a photograph of a sample of DPE and TiD$_2$ after x-ray exposure in the planchet ready for alpha and beta counting. Note also the Swagelok fixture after opening the head.

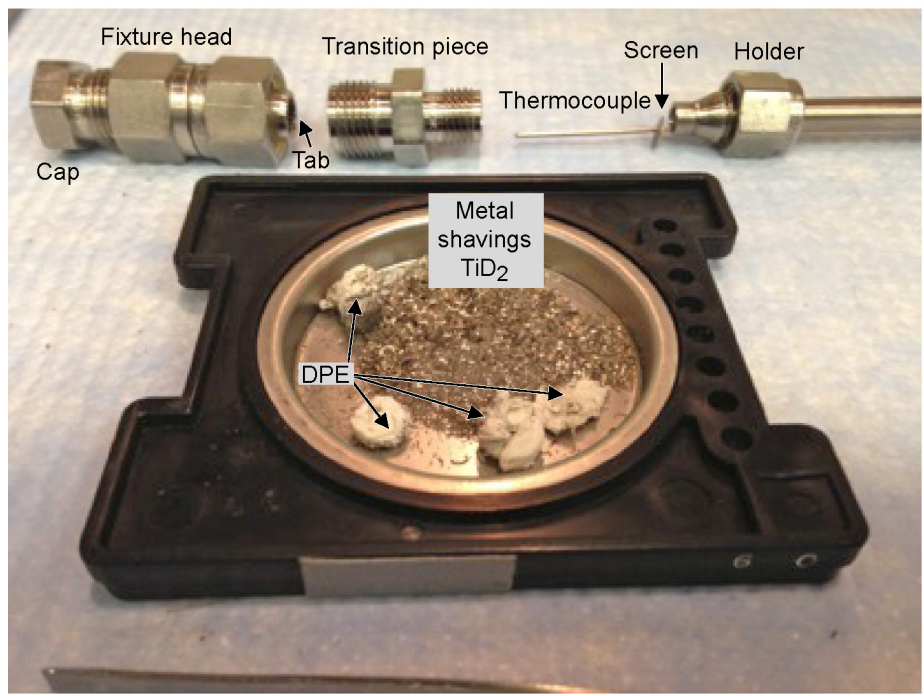

Figure 5.—Photograph of deuterated polyethylene (DPE) plus TiD$_2$ material of SL4 in planchet after x-ray beam exposure prior to alpha and beta scanning.

After alpha and beta counting was complete, all of the samples were also counted with the HPGe detector system. Materials were transferred from the planchet to a plastic sample pan and then placed on top of the HPGe crystal located in the lead cave of the HPGe system.

## 4.1 Posttest Gamma Activity Measurement

None of the samples showed any evidence of gamma activity above the background radioisotopes. Figure 6 shows an example of the gamma spectra collected with one of the samples. The green gamma spectra is a 1-hr count of sample SL16 and the red spectra is a 1-hr count of the empty HPGe cloved cave. Note that both gamma spectra lay directly on top of one another. Peaks labeled in Figure 6 are typical background radioisotopes.

## 4.2 Posttest Alpha and Beta Activity Measurement

A standard 10-min count for alpha activity and then a 10-min count for beta activity were completed using the Canberra (Tennelec) Series 5 XLB—Automatic Low Background Alpha/Beta Counting System for all of the materials after x-ray exposure.

### 4.2.1 Tests With High-Number-Density Deuterium Materials

The test fixtures with DPE, $TiD_2$, and DPE with $TiD_2$ (mixture or layered) were counted for the presence of alpha or beta activity with the Tennelec alpha and beta counting system. A sample is considered to be alpha or beta active when the sample result is above the MDA value.

*Alpha activity*.—There were a few samples that exhibited alpha activity after exposure. Table V shows the samples that showed alpha activity for more than one-half hour after exposure.

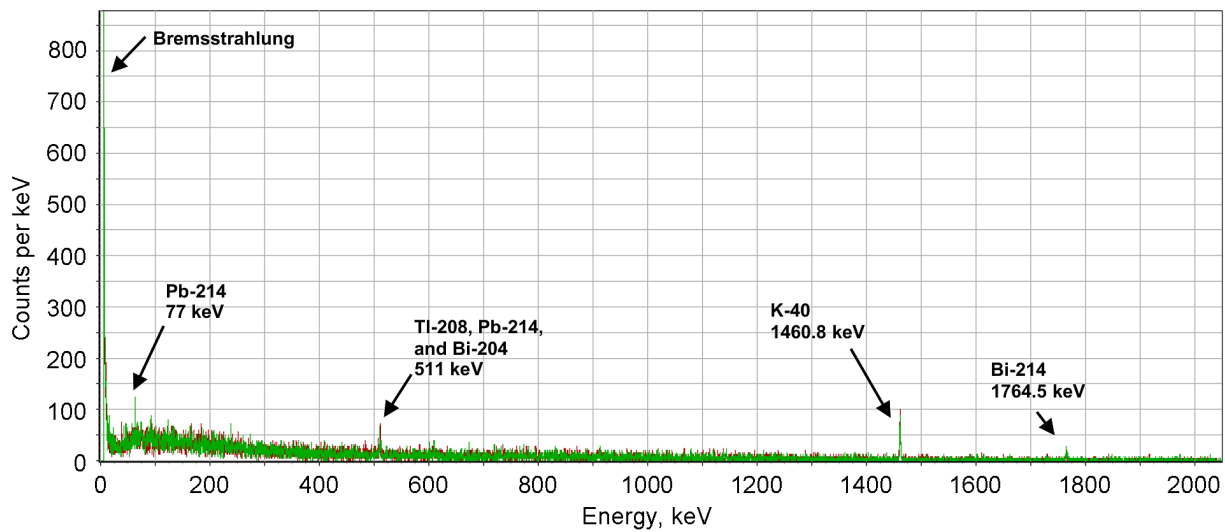

Figure 6.—One-hour gamma spectra of SL16 counted after x-ray exposure. Green spectrum is sample and red spectrum is background empty HPGe closed cave (1-hr count).

TABLE V.—ALPHA AND BETA ACTIVITY OF SAMPLES SHOWING ALPHA ACTIVITY
ABOVE MINIMUM DETECTABLE AMOUNT (MDA) AFTER X-RAY EXPOSURE[a]

| Sample | Materials | Alpha | | | Beta | | |
|---|---|---|---|---|---|---|---|
| | | Alpha, dpm | Uncertainty, dpm | MDA,[b] dpm | Beta, dpm | Uncertainty, dpm | MDA,[b] dpm |
| SL10 | DPE+TiD$_2$ | 4.57 | 1.59 | 2.54 | 14.31 | 4.94 | 6.93 |
| SL10A | DPE+TiD$_2$ | 6.45 | 1.71 | 2.35 | 76.49 | 7.56 | 5.82 |
| SL16 | DPE+TiD$_2$ | 8.44 | 2.1 | 2.54 | 37.95 | 6.47 | 6.93 |
| SL15A | DPE+TiD$_2$ | 6.83 | 1.75 | 2.35 | 9.43 | 5.47 | 5.82 |

[a]Measured in disintegrations per minute (dpm).
[b]MDA is minimum detectable amount.

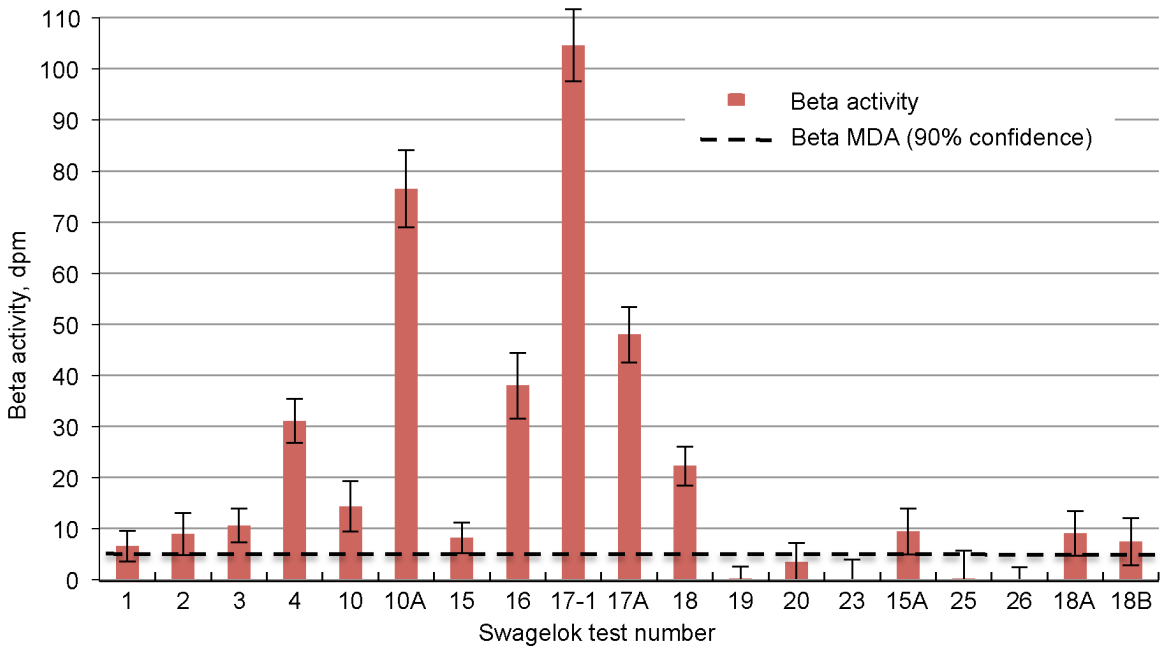

Figure 7.—Beta activities (in disintegrations per minute (dpm)) of deuterated polyethylene (DPE) (SL1 and SL3) and DPE with TiD$_2$ (balance of samples) after x-ray exposure measured by Tennelec alpha and beta counting system compared to that of minimum detectable amount (MDA) (90% confidence).

*Beta activity*.—Figure 7 shows the results of beta counts for the DPE alone (SL1 and SL3) and the remaining test fixtures packed with varying amounts of DPE and TiD$_2$. The DPE alone from the earliest batch of materials received showed activity above their respective MDA (90% confidence level). The beta counts for all of the remaining tests were above their respective MDAs with the exception of SL19, SL23, SL25, and SL26.

It is unclear why SL19, SL20, SL23, SL25, and SL26 did not show beta activity above their MDAs. However, it corresponds with a later shipment from the DPE manufacturer. Work is underway to determine if there were any subtle differences in materials that could have led to activation not being observed.

*Re-exposure*.—Since there was a limited supply of DPE materials because of cost and availability, several of the fixtures were repacked and re-exposed to x-ray photon flux; these are denoted as SL10A, SL17A,[1] SL15A, SL18A, and SL18B. The purpose was to see if the activity would be increased with an additional dosing of x-ray flux. In the case of SL15A the activity of the SL15 was no longer measurable with the Tennelec alpha and beta counting system, but then was reactivated by re-exposure to the x-ray beam. All of the re-exposures resulted in beta activity above MDA. Two (SL10A and SL15A) of the five re-exposures resulted in higher initial alpha and beta activity directly after exposure. The alpha and beta activity results for the initial exposures compared to the re-exposed test fixtures are shown in Table VI. Since SL10A and SL15A contained mixtures of DPE and $TiD_2$, and the rest contained alternating layers of DPE and $TiD_2$, it appears that the mixture made a difference in higher alpha and beta activity.

---

[1]It is noted that SL17A had a small amount of additional material added to fully pack the head.

### 4.2.2 Control Tests

Additional tests were performed as control cases to help better understand the parameter space for what was or was not leading to activation.

*HPE materials*.—The results of the x-ray tests with just HPE material are shown in Table VII. The HPE exposed to x-rays showed no alpha or beta activity above MDA. This finding, when compared with others that exhibited activity, suggests that the addition of high-number-density deuterium is fundamental to obtaining alpha and beta activity.

*No x-ray exposure*.—The alpha and beta activity results for SL61 (vacuum) without x-ray exposure are shown in Table VIII. This test showed that simply loading and unloading the Swagelok fixtures and scanning did not cause the materials to be alpha and beta active (i.e., there did not seem to be a contamination step in the process).

TABLE VI.—COMPARISON OF ALPHA AND BETA ACTIVITY OF SAMPLES
AFTER INITIAL X-RAY EXPOSURE AND RE-EXPOSURE[a]
[Negative activity entries are reported as zero. See Appendix A for more details.]

| Sample | Materials | Alpha | | | Beta | | |
|---|---|---|---|---|---|---|---|
| | | Alpha, dpm | Uncertainty, dpm | MDA,[b] dpm | Beta, dpm | Uncertainty, dpm | MDA,[b] dpm |
| SL10 | DPE+$TiD_2$ mixture | 4.57 | 1.59 | 2.54 | 14.31 | 4.94 | 6.93 |
| SL10A | DPE+$TiD_2$ mixture | 6.45 | 1.71 | 2.35 | 76.49 | 7.56 | 5.82 |
| SL15 | DPE+$TiD_2$ mixture | 2.41 | 1.06 | 2.08 | 4.57 | 3.01 | 4.78 |
| SL15A | DPE+$TiD_2$ mixture | 6.83 | 1.75 | 2.35 | 9.43 | 5.47 | 5.82 |
| SL17 | DPE/$TiD_2$ layers | 2.03 | 0.99 | 2.08 | 104.56 | 7.07 | 4.78 |
| SL17A | DPE/$TiD_2$ layers | 0.0 | 0.53 | 2.35 | 23.08 | 5.41 | 5.82 |
| SL18 | DPE/$TiD_2$ layers | 0.51 | 0.62 | 2.08 | 22.24 | 3.82 | 4.78 |
| SL18A (scan 1) | DPE/$TiD_2$ layers | 1.88 | 1.08 | 2.35 | 3.40 | 4.83 | 5.82 |
| SL18A (scan 2) | DPE/$TiD_2$ layers | 1.50 | 1.01 | 2.35 | 9.03 | 5.02 | 5.82 |
| SL18B | DPE/$TiD_2$ layers | 1.12 | 0.93 | 2.35 | 7.42 | 4.93 | 5.82 |

[a]Measured in disintegrations per minute (dpm).
[b]MDA is minimum detectable amount.

TABLE VII.—ALPHA AND BETA ACTIVITY OF HYDROGEN-BASED POLYETHYLENE (HPE)
MATERIALS AFTER X-RAY EXPOSURE[a]
[Negative activity entries are reported as zero. See Appendix A for more details.]

| Sample | Material | Alpha | | | Beta | | |
|---|---|---|---|---|---|---|---|
| | | Alpha, dpm | Uncertainty, dpm | MDA,[b] dpm | Beta, dpm | Uncertainty, dpm | MDA,[b] dpm |
| SL6 | HPE | 0.69 | 0.81 | 2.54 | 0 | 3.47 | 6.93 |
| SL8 | HPE | 0 | 0.44 | 2.54 | 2.49 | 3.51 | 6.93 |
| SL14 | HPE | 0.21 | 0.65 | 2.54 | 0 | 3.25 | 6.93 |
| SL21 | HPE | 0.35 | 0.76 | 2.35 | 0 | 4.41 | 5.82 |
| SL24 | HPE | 0.73 | 0.85 | 2.35 | 0 | 4.48 | 5.82 |

[a]Measured in disintegrations per minute (dpm).
[b]MDA is minimum detectable amount.

TABLE VIII.—ALPHA AND BETA ACTIVITY OF CONTROL TESTS SL61 AND SL65 BEFORE AND AFTER X-RAY EXPOSURE[a]

[Negative activity entries are reported as zero. See Appendix A for more details.]

| Sample | Alpha | | | Beta | | |
|---|---|---|---|---|---|---|
| | Alpha, dpm | Uncertainty, dpm | MDA,[b] dpm | Beta, dpm | Uncertainty, dpm | MDA,[b] dpm |
| SL61: DPE/TiD$_2$ (vacuum, no x-ray) | | | | | | |
| Before exposure | 0.76 | 0.77 | 2.35 | 0.0 | 3.77 | 5.04 |
| After exposure | 0.38 | 0.66 | 2.35 | 0.0 | 3.79 | 5.04 |
| SL65: Ti (no deuterium loading, with x-ray) | | | | | | |
| Before exposure | 0.0 | 0.54 | 2.3 | 0.0 | 3.79 | 5.04 |
| After exposure | 0 | 0.47 | 2.35 | 3 | 4.67 | 5.82 |

[a]Measured in disintegrations per minute (dpm).
[b]MDA is minimum detectable amount.

*No fuel, with x-ray exposure*.—The alpha and beta activity results for SL65 investigated whether or not the ionization of the titanium materials alone in the x-ray beam could result in activation of the reactant materials. Table VIII shows the alpha and beta activity before and after for these materials. It is clear that the process of simply exposing materials (without deuterium fuel) to the x-ray beam did not cause the materials to become alpha and beta active.

### 4.3 Persistent Beta Activity

Several samples, including SL10A, SL16, and SL17A, showed beta activity after exposure that was over 7 times that of the MDA. Note that SL10A was the re-exposed sample from SL10 and SL17A was the re-exposed sample from SL17. Those samples were selected for subsequent alpha and beta scans to track any long-term beta activity levels.

#### 4.3.1 Sample SL16

Soon after testing SL16, the material was periodically scanned to see how long the activity would persist. Figure 8 shows the beta activity versus scan day, showing that the sample has been beta active for over 12 months since first exposed. Although alpha activities were measured above MDA on the first day, they decayed away soon after the initial alpha and beta scans. Figure 8 shows some variability in the measured beta activity. Some possible reasons are as follows:

(1) For each scan day, the sample was poured from its sealed bottle into a precleaned and prescanned planchet, scans were completed, and then were returned to the bottle. Reloading the sample into the planchet each time likely presented the active material to the detector in differing amounts, resulting in higher or lower recorded counts.

(2) It is also noted that the material from SL16 is rather chunky and is subjected to self-shielding of beta activity. This might account for some measurement variability seen over time.

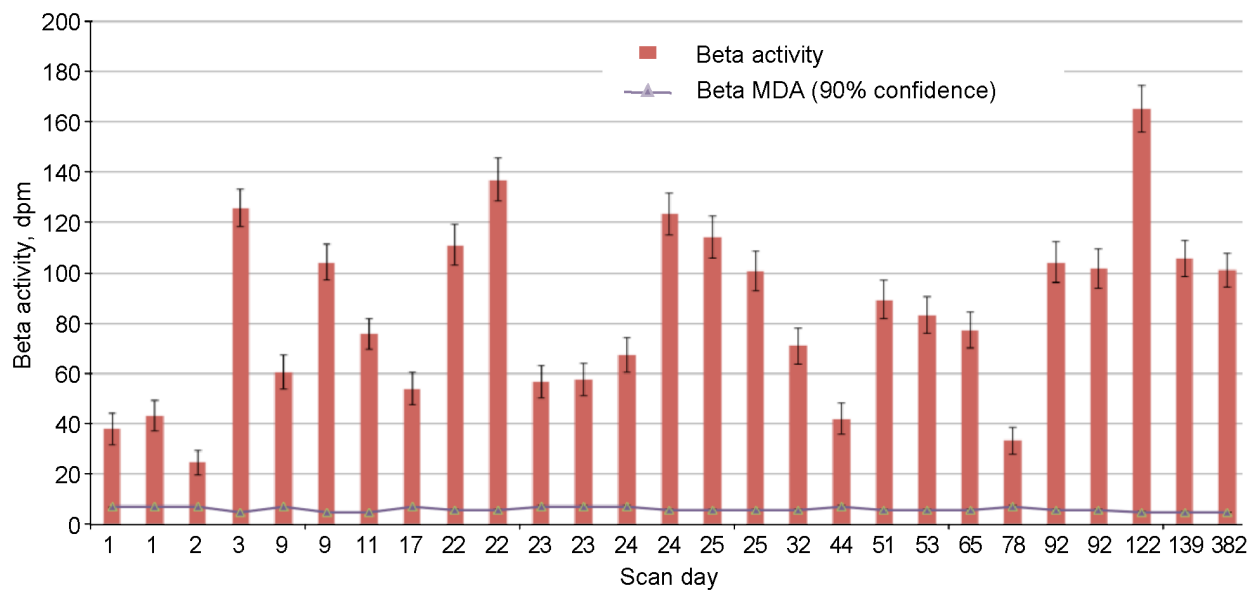

Figure 8.—Beta activity of SL16 measured in disintegrations per minute (dpm) with respect to minimum detectable amount (MDA) for 12+ months after initial exposure to the x-ray beam.

### 4.3.2 Repeat Counting and Statistical Treatment of Samples SL10A, SL16, and SL17A

The alpha and beta activity of SL16 was reassessed approximately 4½ months after initial x-ray beam exposure, and that of SL10A and SL17A was reassessed approximately 3½ months after initial x-ray beam exposure. It is noted that in all cases the alpha activity had gone below MDA. However, in each case the beta activity was persisting, which was the basis for the analyses described in the next several subsections.

#### 4.3.2.1 Multirepeat Beta Scans: ~3 to 4 Months After X-Ray Exposure

Table IX provides the results from a multirepeat scan using the Tennelec machine taken 3 to 4 months after x-ray exposure. Each sample was counted several times in succession (without moving the material within the planchet) to gather enough data to perform a statistical analysis. A tray containing x-rayed sample, referred to as "DPE-TiD$_2$," was examined for beta activity. Sequentially either one or two blank trays were also examined and served as control samples. The objective was to determine if the measurements of the beta activity were different between the two samples (blank tray vs. sample tray) using statistical techniques to a confidence level of 99%. For reference purposes, SL10A, SL16, and SL17A x-ray exposure dates and initial beta scan results are shown in Table X.

The data in Table IX were collected in an interlaced order in two different ways: one way for SL10A and another way for SL16 and SL17A. The data for sample SL10A were collected such that the beta count of a blank tray was recorded, then the first beta count of the DPE-TiD$_2$ sample was recorded, and then the first beta count of another blank tray was collected. This process was repeated until 10 beta-count data points were collected from the DPE-TiD$_2$ sample and 10 from each of the blank trays. The data for samples SL16 and SL17A were collected such that the first beta count of DPE-TiD$_2$ was recorded, and then the first beta count of the blank tray was collected. This process was repeated until 11 beta count data points were collected from the DPE-TiD$_2$ sample and 10 from the blank tray.

TABLE IX.—BETA ACTIVITY OF DPE-TID$_2$ SAMPLES SL10A, SL16, AND SL17A;
RESPECTIVE "CONTROL" BLANK TRAY ACTIVITIES;[a] AND SCAN DATES

| Scan date: Feb. 6, 2015 | | | Scan date: Feb. 1, 2015 | | Scan date: Jan. 31, 2015 | |
|---|---|---|---|---|---|---|
| Blank tray A47, dpm | SL10A, dpm | Blank tray A51, dpm | SL16, dpm | Blank tray for SL16, dpm | SL17A, dpm | Blank tray for SL17A, dpm |
| –0.7 | 27.81 | –1.91 | 50.09 | –0.7 | 33.8 | –1.51 |
| –1.51 | 35.84 | 0.5 | 57.93 | –0.7 | 37.9 | –0.7 |
| 3.31 | 30.62 | –1.91 | 72.78 | –1.51 | 43.1 | –3 |
| –1.91 | 33.03 | –0.3 | 83.22 | 0.9 | 47.9 | –1.61 |
| 1.71 | 35.44 | 0.5 | 94.47 | –1.91 | 49.5 | 1.71 |
| –1.51 | 39.45 | –1.51 | 101.29 | –0.3 | 57.1 | 0.9 |
| 0.9 | 40.66 | 0.1 | 100.09 | –1.1 | 32.2 | –0.3 |
| 0.9 | 39.85 | –1.1 | 92.86 | –0.3 | 63.2 | –2.71 |
| –1.91 | 49.49 | 0.5 | 112.14 | 0.5 | 44.3 | –2.31 |
| –0.7 | 52.3 | –1.1 | 95.67 | –1.91 | 40.7 | –2.71 |
| | | | 105.71 | | 34.2 | |

[a]Measured in disintegrations per minute (dpm).

TABLE X.—REFERENCE INITIAL BETA-SCAN RESULTS[a] AND
X-RAY EXPOSURE DATES FOR SL10A, SL16, AND SL17A

| Fixture | X-ray exposure date | Initial beta activity, dpm | Uncertainty, dpm | MDA,[b] dpm |
|---|---|---|---|---|
| SL16 | Sept. 16, 2014 | 37.95 | 6.47 | 6.93 |
| SL10A | Oct. 13, 2014 | 76.49 | 7.56 | 5.82 |
| SL17A | Oct. 21, 2014 | 47.98 | 6.36 | 5.82 |

[a]Measured in disintegrations per minute (dpm).
[b]MDA is minimum detectable amount.

TABLE XI.—ANDERSON-DARLING
NORMALITY TEST FOR SL10A, SL16, AND SL17A
[Data p-value was greater than alpha of 0.1 for 99% confidence
level—therefore data are treated as normally distributed.]

| Fixture | Normality test Anderson-Darling (Ref. 7) | |
|---|---|---|
| | p-value | Alpha for 99% confidence level |
| SL10A | 0.52 | 0.1 |
| SL16 | 0.166 | 0.1 |
| SL17A | 0.597 | 0.1 |

#### 4.3.2.2 Normality Test

The data were collected for each of the fixtures to determine if it could be treated as normally distributed. The Anderson-Darling test (Ref. 7) was applied at a confidence level of 99%. The computed p-values were all greater than the corresponding alpha value of 0.1 for the 99% confidence level, as shown in Table XI. According to the test, there was insufficient evidence to suggest that the beta counts for SL10A, SL16, and SL17A were not normally distributed. Hence, they were treated as normally distributed for subsequent tests.

#### 4.3.2.2.1 Outlier Tests

A Grubbs test (Ref. 7) was performed on the three data sets to determine if any data set contained a potential outlier. The Grubbs test assumes that if there are any outlier data points, there exists only one.

Either the smallest data point or the largest may be considered an outlier. The test statistic is computed using Equation (1)

$$G = \frac{\max|x_i - \bar{x}|}{s} \quad (1)$$

and is compared against the critical value

$$G_{\text{critical}} = \frac{(n-1)t_{\text{critical}}}{n(n-2+t_{\text{critical}}^2)} \quad (2)$$

where
- $x$     data point
- $s$     standard deviation
- $n$     number of data points
- $t_{\text{critical}}$   critical value of the t-distribution

The test statistic was found not to be larger than the critical value; therefore, there was insufficient evidence to exclude the smallest or largest values from the data sets from further examination. Furthermore, there was no evidence that any DPE-TiD$_2$ data should have been excluded for any other experimental procedural reason. Since neither test statistic for the blank trays was larger than the critical value, there was insufficient evidence to exclude any values from the data sets from further examination.

### 4.3.2.2.2  Means Comparison Tests

Next the means of the "active" samples SL10A, SL16, and SL17A were compared to that of their respective "blank" or "control" planchets to determine what statistical statements could be made of the differences. For the present comparison tests the data sets were not pooled, and the number of degrees of freedom used was 10. The comparison of the means of the beta counts was conducted using a two-sample t-test using a 99% confidence.

### 4.3.2.2.3  T-Test Results

The comparison of the means of the SL10A beta counts was conducted using a two-sample t-test (Ref. 7) with a 99% confidence. The difference of the blank tray A47 and DPE+TiD$_2$ sample mean beta counts was estimated to be 38.45 and resulted in a 99% confidence interval that did not encompass zero (the hypothesized difference between the means). There was sufficient evidence to conclude there was a difference between the mean measurements. The computed p-value was 0.000, meaning that a statement—that the means of the two sample sets were different—would have a 0.000% chance of being incorrect at the 99% confidence interval. Similar results were found when comparing SL10A to the blank tray A51.

When comparing SL16 and SL17A to their respective blank control trays, similar statements could be made at a 99% confidence level. Because the differences were so significant, additional analyses were performed at higher confidence levels. The means of the active samples were found to be different than their respective blanks to greater than a confidence level of 99.999%, corresponding to greater than 4σ.

### 4.3.3  Multirepeat Beta Scans: ~12 months after X-Ray Exposure

Table XII(a) provides the results from a multirepeat scan of DPE-TiD$_2$ samples SL10A, SL16, SL17A using the Tennelec machine taken ~12 months after x-ray exposure. Each sample was counted several times in succession (without moving the material within the planchet). These were collected as sample first, then blank, and that "batch" was repeated 5 times. The blank tray was interspersed and served as control samples to rule out possible contamination concerns. Beta activity for the x-ray-exposed samples ranged from 5 to more than 20 times the MDA, more than 12 months after x-ray exposure.

TABLE XII.—BETA ACTIVITY AND UNCERTAINTIES OF SAMPLES AND
RESPECTIVE "CONTROL" BLANK TRAYS, SCANNED ~12 MONTHS AFTER X-RAY EXPOSURE[a]

(a) Deutrated polyethylene (DPE) plus TiD$_2$ samples SL10A, SL16, and SL17A

| Scan date: Oct. 2, 2015 | | | | Scan date: Oct. 2, 2015 | | | | Scan date: Oct. 5, 2015 | | | |
|---|---|---|---|---|---|---|---|---|---|---|---|
| Blank tray A85, dpm | | SL10A (0.6572g), dpm | | Blank tray A85, dpm | | SL16 (0.7392g), dpm | | Blank tray A85, dpm | | SL17A (1.0678g), dpm | |
| Beta | Unc | Beta | Unc | Beta | Unc | Beta | Unc | Beta | Unc | Beta | Unc |
| 0 | 2.2 | 32.41 | 4.23 | 0 | 2.2 | 99.61 | 6.68 | 1.15 | 2.43 | 20.3 | 3.86 |
| 0.37 | 2.23 | 33.97 | 4.26 | 0 | 2.13 | 113.28 | 7.08 | 0 | 2.2 | 32.41 | 4.26 |
| 0 | 2.13 | 41 | 4.58 | 0 | 2.13 | 100.78 | 6.76 | 1.15 | 2.43 | 20.69 | 3.68 |
| 0 | 2.13 | 37.49 | 4.42 | 0 | 2.136 | 97.26 | 6.63 | 0 | 2.13 | 23.81 | 3.85 |
| 0 | 2.2 | 58.58 | 5.31 | 0 | 1.94 | 101.17 | 6.73 | 0 | 2.2 | 24.2 | 3.83 |

(b) Hydrogen-based polyethylene (HPE) samples SL14, SL21, and SL24

| Scan date: Dec. 10, 2015 | | | | Scan date: Dec. 8, 2015 | | | | Scan date: Dec. 8, 2015 | | | |
|---|---|---|---|---|---|---|---|---|---|---|---|
| Blank tray A90, dpm | | SL14 (0.4545g), dpm | | Blank tray A90, dpm | | SL21 (0.5833g), dpm | | Blank tray A90, dpm | | SL24 (0.6698g), dpm | |
| Beta | Unc | Beta | Unc | Beta | Unc | Beta | Unc | Beta | Unc | Beta | Unc |
| 0 | 1.91 | 1.13 | 2.17 | 0 | 1.95 | 0.35 | 2.03 | 0.74 | 2.14 | 0 | 1.87 |
| 0 | 2.03 | 0 | 1.95 | 0 | 1.99 | 0 | 2.1 | 0 | 1.91 | 0 | 1.91 |
| 0 | 1.91 | 0 | 1.95 | 0 | 1.95 | 0 | 1.95 | 0 | 2.03 | 0 | 1.83 |

[a]Beta minimum detectable amount (MDA): 4.52 disintegrations per minute (dpm), all samples;
Negative beta counts set to zero.

Table XII(b) provides the results from a similar multirepeat scan of HPE "control" samples SL14, SL21, and SL24 using the Tennelec machine taken ~12 months after x-ray exposure. Each sample was counted several times in succession (without moving the material within the planchet). These were collected as blank first, then sample, and that "batch" was repeated three times. The blank tray was examined and served as control samples to rule out possible contamination concerns. No beta activity was observed.

### 4.3.4 Beta Scintillation Study of Baseline As-Received DPE and Select Active Samples After X-Ray Exposure

Beta scintillation assessments used the LS–6500 beta scintillator for both baseline, or as-received, DPE materials, and portions of the SL10A, SL16, and SL17A materials to determine if there were persistent beta emissions.

#### 4.3.4.1 Beta Scintillation Study: Apparatus and Procedures

The LS–6500 automatically integrates the counts in the tritium, $^{14}$C, and $^{32}$P beta energy bands. The system is calibrated periodically according to Beckman specifications to confirm that the machine is properly counting the corresponding standard.

##### 4.3.4.1.1 Baseline As-Received DPE Material: External Laboratory

Two pieces of as-received DPE materials denoted Sample J and BC 0341634 were taken from two batches of as-received materials including those that showed beta activity after exposure. These were counted with blank vials of Scintiverse scintillation cocktail at a co-author's laboratory external to NASA Glenn.

##### 4.3.4.1.2 SL17A: External Laboratory

A portion of SL17A sample material was sent to an external laboratory for beta scintillation scans. Another portion was maintained at NASA Glenn for internal scans as will be described below. The portion of sample of SL17A sent to the external manufacturer was divided into the DPE and TiD$_2$ components, and each sample was placed into a vial with 10 ml of Scintiverse scintillation cocktail. These two segregated samples were counted along with a "blank" of the same amount of scintillation fluid for reference purposes. It was counted for various intervals over the period of time from October 27, 2014, through February 8, 2015.

#### 4.3.4.1.3 SL10A, SL16, and SL17A: NASA Glenn Laboratory

A portion of each of the samples exposed at SL10A, SL16, and SL17A (containing both TiD$_2$ and DPE) were placed into 10 ml of Ecoscint fluid for beta scintillation scans. These three samples were counted along with a "blank" of the same amount of scintillation fluid for reference purposes. The scans were performed at least 12 months after exposure to investigate the presence beta energies in the tritium channel.

### 4.3.4.2 Beta Scintillation Results

#### 4.3.4.2.1 Baseline As-Received DPE Material

Table XIII presents results for two pieces of as-received DPE materials denoted Sample J and BC 0341634, taken from two batches of as-received materials including those that showed beta activity after exposure. These as-received materials were counted with blank vials of Scintiverse scintillation cocktail. One can see that the activity in the tritium energy band is consistent with the blank vial. So it appears that the material exhibiting energies in the tritium band were not there in the starter materials. The table also shows the LumEx value (percent measure of chemoluminescence) and the H-# (measure of quench or sample photon self-absorption).

#### 4.3.4.2.2 SL17A: External Laboratory

A portion of the SL17A was separated into its two major components: DPE and TiD$_2$. These elements were scanned separately in the beta scintillator at the external lab. Figure 9 shows that the DPE from SL17A is showing activity in the tritium channels greater than background, approaching about 2 times background. For the scan results presented, the following scan parameters were observed: (1) LumEx was less than 1, showing no chemoluminescence present; (2) H-# was less than the desired threshold of 120, indicating sample generated light was not being absorbed and lowering the counts; and (3) The 60-min counts consistently give an error <5 percent.

TABLE XIII.—BETA SCINTILLATION RESULTS OF BASELINE AS-RECEIVED DEUTERATED POLYETHYLENE (DPE) MATERIALS

| Date | Sample | Count duration, min | Tritium band,[a] cpm | LumEx | H-# |
|---|---|---|---|---|---|
| Dec. 3, 2014 | DPE Sample J | 240 | 53.30 | 1.73 | 81.7 |
| | Blank | 240 | 51.29 | 0.41 | 71.7 |
| Feb. 13, 2015 | DPE Sample J | 10 | 49.50 | 4.32 | 138.6 |
| | Blank | 10 | 52.8 | 0.59 | 72.3 |
| Feb. 12, 2015 | DPE Sample Ref BC #0341634 | 60 | 58.28 | 5.17 | 84.8 |
| | Blank | 60 | 49.83 | 1.01 | 71.2 |
| Feb. 12, 2015 | DPE Sample Ref BC #0341634 | 60 | 53.93 | 3.84 | 85.3 |
| | Blank | 60 | 51.5 | 0.85 | 72.4 |
| Feb. 13, 2015 | DPE Sample Ref BC #0341634 | 10 | 52.5 | 1.79 | 86.1 |
| | Blank | 10 | 52.8 | 0.59 | 72.3 |

[a]Measured in counts per minute (cpm).

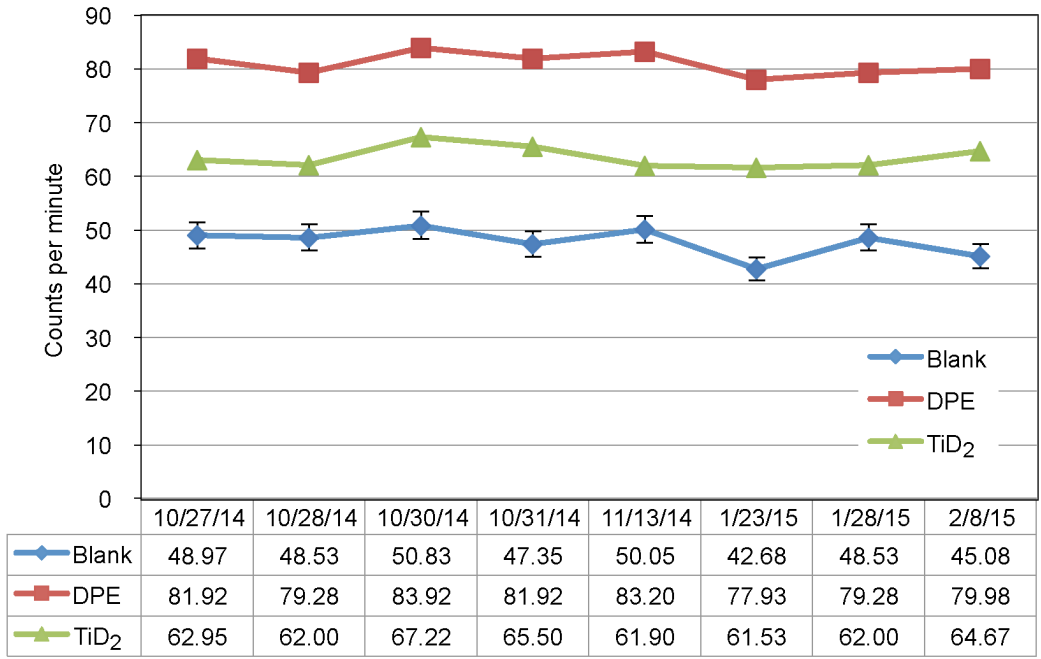

Figure 9.—Beta scintillation data (1 to 18 keV spectral band, ±5 percent uncertainty) versus run number for DPE, TiD$_2$ (from SL17A), and blank vials.

*Comparison of Means to Blank*.—A statistical means test was performed comparing the beta scintillation results of the DPE and TiD$_2$ samples from SL17A to the blanks. The means of the active samples were found to be different from the blank to greater than a confidence level of 99.999%, corresponding to greater than 4σ.

#### 4.3.4.2.3   SL10A, SL16, and SL17A: NASA Glenn Laboratory

Table XIV shows that the SL10A sample is showing activity in the tritium channel about 2 to 5 times that of the blank vial, depending on scan date. SL16 is showing beta activity in the tritium channel 6 times that of the blank vial. Sample 17A (mixture of both DPE and TiD$_2$) is showing activity in the tritium channels 2 to 2.5 times that of the blank vial. For the scan results presented, the 60-min counts consistently give an error <6 percent. SL17A has been counted on two different Beckman LS–6500 liquid scintillator spectrometers, one using the Ecoscinct cocktail and the other Scintiverse. Both spectrometers have given similar results multiple times for many months.

Figure 10 shows beta scintillation counts versus energy (keV) for SL10A from the LS–6500 1-hr-long scans. Plotted also for reference are the results for tritium calibration standard. Tritium has a beta energy peak at 5.7 keV and a maximum beta energy at 18.6 keV (often referred to at the end-point beta energy). These 1-min scan results were scaled downward by a factor of about 7:1 to place them on an equal basis, for comparison purposes. SL10A shows evidence of beta activity consistent with tritium, as detected by scintillation spectroscopy. It is noted that the peak counts versus energy is shifted slightly below that of the tritium standard. The reason for this is unclear, but may have been caused by energy down-scattering of the betas by the TiD$_2$ and DPE matrix from which they are emanating, combined with low-count statistics. Previous liquid scintillator spectroscopic analysis of the un-irradiated TiD$_2$ and/or DPE show no evidence of tritium. The observed beta activity has continued for over 12 months, and there are no other likely beta emitters with end-points below 18 keV and half-lives in years other than tritium. This indicates that the x-ray irradiation of the TiD$_2$+DPE matrix resulted in the production of tritium by an unexpected nuclear effect.

TABLE XIV.—BETA SCINTILLATION RESULTS OF DEUTERATED POLYETHYLENE (DPE) PLUS TiD$_2$
SAMPLES SL10A, SL16, AND SL17A SCANNED ~12 MONTHS AFTER X-RAY EXPOSURE

| Sample mass, g | X-ray exposure date | Liquid scintillator scan date | Sample | Count duration, min | Tritium band,[a] cpm | LumEx | H# |
|---|---|---|---|---|---|---|---|
| SL10A 0.29874 | Oct. 13, 2014 | Oct. 8, 2015 | Blank | 60 | 27.45 | 0.02 | 26.1 |
| | | | DPE+TiD$_2$ | 60 | 144.37 | 0.10 | 55.7 |
| | | Dec. 8, 2015 | Blank | 60 | 24.82 | 0.04 | 26.4 |
| | | | DPE+TiD$_2$ | 60 | 57.87 | 0.06 | 72.2 |
| SL16 0.26836 | Sept. 16, 2014 | Dec. 8, 2015 | Blank | 60 | 25.68 | 0.02 | 26.2 |
| | | | DPE+TiD$_2$ | 60 | 172.43 | 0.07 | 74.0 |
| SL17A 0.10648 | Oct. 21, 2014 | Oct. 8, 2015 | Blank | 60 | 20.98 | 0.15 | 32.3 |
| | | | DPE/TiD$_2$ | 60 | 42.05 | 0.11 | 53.5 |
| | | Dec. 8, 2015 | Blank | 60 | 19.42 | 0.05 | 38.6 |
| | | | DPE/TiD$_2$ | 60 | 49.08 | 0.03 | 72.4 |

[a]Measured in counts per minute (cpm).

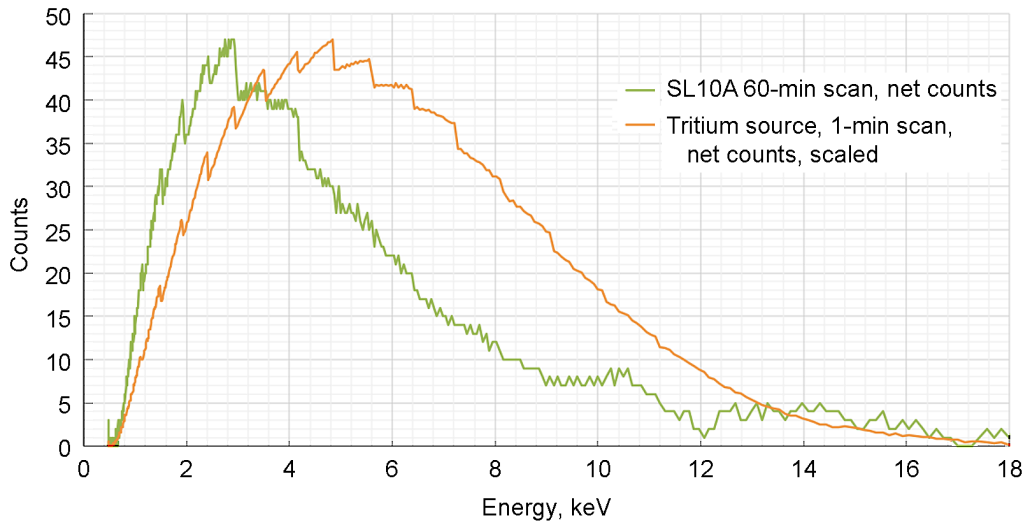

Figure 10.—Beta scintillation data spectral band, (±5 percent uncertainty) for SL10A plotted with results for the tritium-calibration standard (scaled by 0.1451), for comparison purposes. Both data sets have backgrounds subtracted.

## 5.0 Summary of Results

A series of experiments were conducted exposing various deuterated and nondeuterated materials to x-ray irradiation at various energies ranging from 65 to 280 kV, with prescribed currents impinging upon the tungsten braking target from 500 to 9000 μA for various total exposure times ranging from 55 to 280 min. Swagelok stainless steel fixtures were used to hold the various materials under either vacuum or in a few cases low (<350 psia (2.4 MPa)) deuterium gas pressures. Different materials were investigated including deuterated polyethylene (DPE), titanium deuteride (TiD$_2$) and mixtures thereof. For control purposes, hydrogen-based polyethylene (HPE) was also examined, as were unloaded Ti powders. Scans of as-received materials were completed to document alpha and beta activity rates before exposure. Materials were shown to have no alpha or beta activity above the minimum detectable amount (MDA) before exposing them to the x-ray beam.

*Control tests*.—Five different tests were performed with HPE (control, no deuterium fuel) samples that showed posttest alpha or beta activity that was not above the MDA, after exposing to the noted x-ray beam protocol. In another control test, a combination of DPE and TiD$_2$ particles were loaded into the Swagelok fixture and placed in the x-ray laboratory, but the beam was not energized. As expected, the test showed alpha or beta activity that was not above MDA. Another test examined if placing titanium powders (without deuterium loading) in the Swagelok fixture and exposing to the ionizing x-ray beam would result in activation. That test also showed alpha or beta activity that was not above MDA.

***DPE and TiD$_2$ tests***.—Two tests performed with DPE alone showed beta activity above background. TiD$_2$ mixed with DPE samples were the most active in regards to beta activity.

Fourteen tests out of 19 total runs in this test sequence with either DPE or DPE with TiD$_2$ were beta activated. Some samples exhibited alphas, which decayed below MDA in approximately an hour following x-ray exposure. Several of the samples with DPE and TiD$_2$ showed persistent beta activity. Several of the samples (SL10A, SL16, and SL17A) showed beta activity above background with a greater than 4$\sigma$ confidence level for months after exposure. Portions of SL10A, SL16, and SL17A were scanned using a beta scintillator and found to have beta counts in the tritium energy band, continuing without noticeable decay for over 12 months. Beta scintillation investigation of as-received materials (before x-ray exposure) showed no beta counts in the tritium energy band, indicating the beta emitters were not in the starting materials.

As noted, in this test sequence, five of the fixtures containing both DPE and TiD$_2$ were not beta activated after x-ray exposure. The underlying reason is not presently clear but is currently under investigation by examining whether it is related to material exposure time underneath the x-ray beam, insufficient loading of deuterium in the material, and/or other possible material factors.

# Appendix A.—Canberra (Tennelec) Series 5 XLB Automatic Low Background Alpha and Beta Counting System

In this study, as-received and postexposed materials were scanned with a Canberra (Tennelec) Series 5 XLB Automatic Low Background Alpha/Beta Counting System gas proportional counter (Ref. 8). The standard gas flow detector of the Series 5 family of systems incorporates a high-performance pancake-style 5.7-cm- (2.25-in.-) diameter detector. The entrance window of the detector is made of thin Mylar to provide the highest counting efficiency and the lowest alpha background of any counter. The standard detector window has an area density of 80 μg/cm². The Series 5 incorporates technology to reduce system background. Using an improved guard detector, the system sensitivity for a high-energy, cosmic background is increased, enabling the anticoincidence circuitry to detect and reject more spurious background events. The unit uses a graded shielding system of 10 cm (4 in.) of custom molded lead surrounding the detector. The beta background for the Series 5 can be as low as 0.5 counts per minute.

The electronics package of the Series 5 family of counting systems provides for control and several user-defined counting approaches. The approach used for this study was to count using alpha-then-beta (ATB) with the "background subtraction" to assure accurate results. Ten-minute count times were used for each of the alpha and alpha/beta intervals. The following sections summarize how the machine evaluates efficiency, minimum detectable amount (MDA), and finally, alpha and beta counts (Ref. 9).

## A.1    Efficiency Description

The efficiency of this system is determined semi-annually and at any time where the quench gas is changed using two calibrated sources ($^{99}$Tc for betas and $^{239}$Pu for alphas). The machine efficiency ($\varepsilon$, roughly 26%) was determined from the following equation:

$$\varepsilon = \frac{1}{N} \cdot \sum_{i=1}^{N} \varepsilon_i = \frac{1}{N} \cdot \sum_{i=1}^{N} \frac{R_{i_{calc}}}{E} \qquad (3)$$

where
- $\varepsilon_i$      efficiency determined from the i$^{th}$ observation
- $R_{i_{calc}}$      calculated rate count
- $N$      total number of times that sample is counted
- $i$      index
- $E$      emission rate of the calibration standard, given by the equation

$$E = E_0 \cdot e^{-\ln(2) \cdot \frac{\Delta T}{T_{1/2}}} \qquad (4)$$

where
- $E_0$      emission rate of the calibration standard on the calibration date
- $\Delta T$      lapsed time between the calibration and the count acquisition
- $T_{1/2}$      half-life of the calibration standard

The uncertainty of the efficiency value ($\sigma_\varepsilon$) was determined from the following equation:

$$\sigma_\varepsilon = \sqrt{\frac{\sum_{i=1}^{N}(\varepsilon_i - \varepsilon)^2}{N-1}} \qquad (5)$$

where
ε    average efficiency
$\varepsilon_i$    individual efficiency value

## A.2   Minimum Detectable Amount (MDA) Description

The MDA of activity is the lowest activity discernable from no activity. The MDA is determined quarterly and is changed if periodic checks show a change in the measured background or efficiency. The MDA uses a 90% confidence error band in the estimate; therefore, no additional error bars are shown in the graphs and tables in the report. The MDA was estimated from the measured background count rate using

$$MDA = \frac{L_{D_{Rate}}}{\varepsilon \cdot S} \cdot F_{AC} \tag{6}$$

where
ε    efficiency (determined approximately twice yearly or when quench gas bottle is changed using 20 each counts of calibrated sources)
$S$    sample size in selected units. For certain sample types (e.g., smears), $S = 1$
$F_{AC}$    unit conversion factor to change reported activity from disintegrations per minute (dpm), if necessary
$L_{D_{Rate}}$    detection limit (in units of rate; e.g., dpm), which is given by

$$L_{D_{Rate}} = \frac{k^2}{T_S} + 2 \cdot L_{C_{Rate}} \tag{7}$$

where
$L_{C_{Rate}}$    critical limit, which is the minimum number of counts required to declare detection from the source and is given by the equation:

$$L_{C_{Rate}} = k \cdot \sqrt{\frac{R_B}{T_S} + \sigma_{R_B}^2} \tag{8}$$

where
$k$    coverage factor (1.645 for 90% confidence level)
$R_B$    reported average background count rate (determined quarterly by the average of 20 blank planchet counts)
$T_S$    sample count time in minutes
$\sigma_{R_B}$    uncertainty of system background count rate, given by the equation:

$$\sigma_{R_B} = \sqrt{\frac{\sum_{i=1}^{N}(R_{Bi} - R_B)^2}{N-1}} \tag{9}$$

where

$R_{B_i}$  individual calculated rate count

Combining the above,

$$MDA = \frac{\left[\dfrac{k^2}{T_S} + 2k \cdot \sqrt{\dfrac{R_B}{T_S} + \sigma_{R_B}^2}\right]}{\varepsilon \cdot S} \cdot F_{AC} \tag{10}$$

## A.3   Counting an Unknown Sample

We have conservatively reported measurements in disintegrations per minute, with a calculated MDA, based upon the measured efficiencies and calibrations using the $^{99}$Tc beta energy of 297.7 keV and the $^{239}$Pu alpha energy of 5.6 MeV. However, since this is an exploratory study where we have an activated sample with unknown radioisotopes, it is unlikely we have either betas or alphas of these specific average energies. There are additional losses due to the sample geometry relative to the calibration geometry and sample self-absorption. Consequently, our calculated dpm likely differs from an ideal dpm, based on these noted parameters. By running a blank as a background with each sample (see Sec. 4.2.2), there is a higher sensitivity than the MDA alone would provide.

## A.4   Alpha and Beta Activity Measurements

The reported alpha count was determined from a single measurement. Since the alpha count was not quantified simultaneously with the beta count, no spillover (spilldown) correction was needed.

### A.4.1   Alpha Count (Using Alpha-Then-Beta (ATB) Mode With Background Subtraction)

$$R_\alpha = R_{\alpha_{gross}} - R_{B\alpha} \tag{11}$$

where
$R_\alpha$  alpha count rate
$R_{\alpha_{gross}}$  the gross (alpha only) count rate obtained during the (alpha-only) count mode
$R_{B\alpha}$  the alpha (system) background count rate for the alpha-only mode of the ATB count

### A.4.2   Alpha Count Uncertainty

$$\sigma_{R_\alpha} = \sqrt{\frac{R_{\alpha_{gross_1}}}{T_1} + \sigma_{R_{B\alpha}}^2} \tag{12}$$

where
$R_{\alpha_{gross_1}}$  the gross (alpha-only) count rate obtained during the (alpha-only) count mode of this one measurement
$T_1$  the count time of this one measurement
$\sigma_{R_{B\alpha}}$  the uncertainty in background count rate (90% confidence level) in the alpha-only mode of the ATB count

### A.4.3 Beta Count (Using ATB Mode)

Each reported beta count was determined from a single measurement. Since the beta count was not quantified simultaneously with as the alpha count, no spillover (spillup) correction was needed. However, the Tennelec alpha and beta counting system determined the beta count through subtraction of the gross (alpha-only) count rate from the gross (alpha + beta) count rate.

$$R_\beta = R_{\alpha+\beta} - R_\alpha - R_{B\,\beta} \tag{13}$$

where

$R_\beta$   beta count rate
$R_{\alpha+\beta}$   gross (alpha + beta) count rate obtained during the (alpha + beta) count mode
$R_\alpha$   gross (alpha-only) count rate obtained during the (alpha-only) count mode
$R_{B\,\beta}$   derived beta system background count rate for the ATB mode

Note: For very low counts, this has the potential to return negative beta count values if the emission of alpha decays during the (alpha-only) count/(alpha + beta) count sequence.

### A.4.4 Beta Count Uncertainty

$$\sigma_{R_\beta} = \sqrt{\left(\frac{R_{\alpha+\beta}}{T}\right) + \frac{R_\alpha}{T} + \sigma^2_{R_{B_{\alpha+\beta}}} + \sigma^2_{R_{B\alpha}}} \tag{14}$$

where

$\sigma_{R_\beta}$   uncertainty in the β count rate of the ATB count mode
$\sigma_{R_{B_{\alpha+\beta}}}$   uncertainty in the (alpha + beta) system background count rate for the (alpha + beta) mode of the ATB count mode.

## Acknowledgments

The authors gratefully acknowledge the assistance of Dr. Kathy Chuang and Mr. Dan Scheiman in performing the infrared spectroscopy of the deuterated polyethylene in measuring C-D to C-H material bond ratios. We also acknowledge Richard Rauser, operator of the microfocus x-ray beam, and Radiation Safety Office personnel, Chris Blasio, and Rod Case, without whose assistance this work would not have been possible. We are also grateful for Gus Fralick's comments on the manuscript and Laura Becker's patient editorial attention. This work was supported by NASA's Planetary Science Division, Science Mission Directorate.